\documentclass[a4paper,11pt]{article}
\usepackage[dvips]{color,graphicx}
\usepackage{stmaryrd}
\def\dprod{\displaystyle\prod}

\def\dsum{\displaystyle\sum}

\def\bt{\mathbf{t}}
\def\Tr{\mathrm{Tr}}
\def\Re{\mathrm{Re}}
\def\Im{\mathrm{Im}}

\usepackage{graphicx,amssymb,amsmath,bm,latexsym}

\allowdisplaybreaks

\begin{document}
\begin{center}
{\Large\bf $W_{1+\infty}$ constraints for the hermitian one-matrix model}\vskip .2in
{\large Rui Wang$^{a}$, Ke Wu$^{a}$, Zhao-Wen Yan$^{b}$, Chun-Hong Zhang$^{c}$,
Wei-Zhong Zhao$^{a}$\footnote{Corresponding author: zhaowz@cnu.edu.cn}} \vskip .2in
$^a${\em School of Mathematical Sciences, Capital Normal University,
Beijing 100048, China} \\
$^b${\em School of Mathematical Sciences, Inner Mongolia University, Hohhot
010021, China}\\
$^c${\em School of Mathematics and Statistics, North China University of Water Resources and Electric Power,
Zhengzhou 450046, China}\\

\begin{abstract}
We construct the multi-variable realizations of the $W_{1+\infty}$ algebra
such that they lead to the $W_{1+\infty}$ $n$-algebra.
Based on our realizations of the $W_{1+\infty}$ algebra,
we derive the $W_{1+\infty}$ constraints for the hermitian one-matrix model.
The constraint operators yield not only the $W_{1+\infty}$ algebra
but also the closed $W_{1+\infty}$ $n$-algebra.
\end{abstract}

\end{center}



\section{Introduction}

The Virasoro constraints for the matrix models have attracted remarkable attention \cite{Mironov}-\cite{Dijkgraaf}.
Since the $W_{1+\infty}$ algebra can be generated by the higher order differential operators
with respect to the eigenvalues of the matrices,
an approach to derive a large class of constraint equations
for matrix models at finite $N$ was proposed in Ref.\cite{Itoyama}.
These constraints are associated with the higher order
differential operators of $W_{1+\infty}$ algebra, where the well-known Virasoro constraints
are associated with the first order differential operators.
However, it seems rather nontrivial to write down the constraints explicitly.
The Ding-Iohara-Miki (DIM) algebra is a quantum deformation of the toroidal algebra with
two central charges \cite{Iohara}-\cite{Awata},
which has attracted much interest from physical and mathematical points of view.
It was found that the Ward identities in the network matrix models can be described in terms of this symmetry \cite{Zenkevich, ZenkevichJHEP}.

The elliptic generalization of hermitian matrix model is known to be associated with the
$4d$ $\mathcal{N}=1$ $U(N)$ gauge theory on $S^3\times S^1$ \cite{Gadde}-\cite{Nedelin}.
The $q$-Virasoro constraints for this matrix model have been derived by the insertion of
the $q$-Virasoro generators under the contour integral \cite{Nedelin},
where the $q$-Virasoro generators are constructed in terms of $q$-derivatives within the $q$-calculus
and the corresponding $q$-Virasoro algebra is a special case of a more general elliptic deformation
of the Virasoro algebra \cite{Shiraishi}.
Since 3-algebra has recently been found useful in the Bagger-Lambert-Gustavsson (BLG)
theory of M2-branes \cite{BL2007, Gustavsson},
the applications of $n$-algebra have aroused much interest \cite{Izquierdo}-\cite{Arvanitakis}.
More recently it was found that there are the generalized $q$-$W_{\infty}$ constraints
for the elliptic matrix model \cite{Wang}.
Although these constraint operators do not yield the closed algebras,
by applying the strategy of carrying out the action of the
operators on the partition function as done in Ref.\cite{Nedelin},
the ($n$-)commutators of the constraint operators lead to the generalized
$q$-$W_{\infty}$ algebra and $n$-algebra, respectively.
For the elliptic matrix model, its partition function also satisfies
the constraints from the elliptic DIM algebra \cite{Zenkevich}.
In this letter,  we focus on the hermitian one-matrix model and derive its $W_{1+\infty}$
constraints. We show that the derived constraint operators yield the $W_{1+\infty}$ $n$-algebra.

\section{ The multi-variable realizations of $W_{1+\infty}$
algebra and its $n$-algebra}
Let us first recall the $W_{1+\infty}$ algebra \cite{Cappelli}
\begin{eqnarray}\label{walgebra}
[W_{m_1}^{r_1}, W_{m_2}^{r_2}]
=(\dsum_{k=0}^{r_1-1}C_{r_1-1}^{k}A_{m_2+r_2-1}^{k}
-\dsum_{k=0}^{r_2-1}C_{r_2-1}^{k}
A_{m_1+r_1-1}^{k})W_{m_1+m_2}^{r_1+r_2-1-k},
\end{eqnarray}
where
$A_n^{k}=\left\{\begin{array}{cc}
n(n-1)\cdots(n-k+1),& k\leqslant n,\\
0,                  &k>n,\end{array}\right. $
$C_n^{k}=\frac{n(n-1)\cdots(n-k+1)}{k !}$.

Its single variable realization is given by
\begin{eqnarray}\label{eq:walgoperator}
W_m^r=z^{m+r-1}\frac{d^{r-1}}{d z^{r-1}}, \ \ r\in \mathbb{Z}_+, m \in\mathbb{Z},
\end{eqnarray}
which not only yields (\ref{walgebra}), but also leads to
the $W_{1+\infty}$ $n$-algebra \cite{Zhang}
\begin{eqnarray}\label{Wnalgebra}
[W_{m_1}^{r_1}, W_{m_2}^{r_2}, \ldots, W_{m_n}^{r_n}]&:=&\epsilon_{1 2 \cdots n}^{i_1 i_2 \cdots i_n}
W_{m_{i_1}}^{r_{i_1}}W_{m_{i_2}}^{r_{i_2}}\cdots W_{m_{i_n}}^{r_{i_n}}\nonumber\\
&=&\epsilon_{1 2 \cdots n}^{i_1 i_2 \cdots i_n} \dsum_{\alpha_1=0}^{\beta_1}
\dsum_{\alpha_2=0}^{\beta_2}
\cdots \dsum_{\alpha_{n-1}=0}^{\beta_{n-1}}
C_{\beta_1}^{\alpha_1}C_{\beta_2}^{\alpha_2}\cdots C_{\beta_{n-1}}^{\alpha_{n-1}}
\cdot A_{m_{i_2}+r_{i_2}-1}^{\alpha_1}\nonumber\\
&&A_{m_{i_3}+r_{i_3}-1}^{\alpha_2}\cdots A_{m_{i_n}+r_{i_n}-1}^{\alpha_{n-1}}
W_{m_1+m_2+\cdots +m_{n}}^{r_1+\cdots+r_n-(n-1)-\alpha_1-\cdots-\alpha_{n-1}},
\end{eqnarray}
where
$
\epsilon _{j_{1}\cdots j_{p}}^{i_{1}\cdots i_{p}}=\det \left(
\begin{array}{ccc}
\delta _{j_{1}}^{i_{1}} & \cdots & \delta _{j_{p}}^{i_{1}} \\
\vdots &  & \vdots \\
\delta _{j_{1}}^{i_{p}} & \cdots & \delta _{j_{p}}^{i_{p}}%
\end{array}%
\right)
$
and
$\beta_k=\left\{
  \begin{array}{cc}
    r_{i_1}-1,& k=1, \\
\dsum_{j=1}^k r_{i_j}-k-\dsum_{i=1}^{k-1}\alpha_i,   & 2\leqslant k\leqslant n-1.\\
  \end{array}\right.$

Since the associativity of the product of the operators (\ref{eq:walgoperator}) holds,
the $n$-algebra (\ref{Wnalgebra}) with $n$ even satisfies
the generalized Jacobi identity (GJI) \cite{Izquierdo}
\begin{equation}\label{GJI}
\epsilon _{1\cdots (2n-1)}^{i_{1}\cdots i_{2n-1}}
[[A_{i_{1}},A_{i_{2}},\cdots ,A_{i_{n}}],A_{i_{n+1}},\cdots ,A_{i_{2n-1}}]=0.
\end{equation}
When $n$ is odd, it satisfies the generalized
Bremner identity (GBI) \cite{CJM09, Weingart}
\begin{eqnarray}\label{GBI}
&&\epsilon _{1\cdots (3n-3)}^{i_{1}\cdots i_{3n-3}}
[[A,B_{i_{1}},\cdots ,B_{i_{n-1}}],[B_{i_{n}},\cdots
,B_{i_{2n-1}}],B_{i_{2n}},\cdots ,B_{i_{3n-3}}]  \notag\\
&&=\epsilon _{1\cdots (3n-3)}^{i_{1}\cdots i_{3n-3}}
[[A,[B_{i_{1}},\cdots ,B_{i_{n}}],B_{i_{n+1}},\cdots
,B_{i_{2n-2}}],B_{i_{2n-1}},\cdots ,B_{i_{3n-3}}].
\end{eqnarray}

Hence the $W_{1+\infty}$ $n$-algebra with $n$ even
is a generalized Lie algebra (or higher order Lie algebra).
A remarkable property of (\ref{Wnalgebra}) is that there are the following subalgebras
\begin{eqnarray}\label{sub2nalgebra}
[W_{m_1}^{n+1}, W_{m_2}^{n+1}, \ldots , W_{m_{2n}}^{n+1}]=\dprod_{1\leqslant j<k\leqslant 2n}(m_k-m_j)
W_{m_1+m_2+\cdots +m_{2n}}^{n+1},
\end{eqnarray}
and
\begin{eqnarray}\label{nullsubalg}
[W_{m_1}^{n+1}, \ldots ,W_{m_{2n+1}}^{n+1}]=0.
\end{eqnarray}

A well-known multi-variable realization of (\ref{walgebra}) is
\begin{eqnarray}\label{mulwalgoperator1}
\bar W_m^r&=&\sum\limits_{i=1}^Nz_i^{m+r-1}\frac{\partial^{r-1}}{\partial z_i^{r-1}},
\ \ r\in \mathbb{Z}_+, m \in\mathbb{Z}.
\end{eqnarray}
However the generators (\ref{mulwalgoperator1}) do not yield
the nontrivial $n$-algebra except for the null $(2nN+1)$-algebra \cite{Zhang}
\begin{eqnarray}\label{conclusion}
[\bar W_{m_1}^{n+1}, \bar W_{m_2}^{n+1}, \cdots, \bar W_{m_{2nN+1}}^{n+1}]=0.
\end{eqnarray}
Note that it is not only determined by the superindex of the generators,
but also the number of variables $N$.

In order to construct the multi-variable realization of $W_{1+\infty}$ $n$-algebra,
let us introduce the Euler operator
\begin{equation}
O_E=\sum_{i=1}^{N}z_i\frac{\partial}{\partial z_i}
\end{equation}
and the Lassalle operators \cite{Lassalle}
\begin{eqnarray}
&&O_L^A=\sum_{i=1}^{N}\frac{\partial^2}{\partial z_i^2}-\frac{2}{\alpha}\sum_{i=1}^{N}\sum_{j\neq i}\frac{1}{z_i-z_j}\frac{\partial}{\partial z_i},\label{LA}\\
&&O_L^B=\sum_{i=1}^{N}(\frac{\partial^2}{\partial z_i^2}+\frac{\beta}{z_i}\frac{\partial}{\partial z_i})+\gamma\sum_{i=1}^{N}\sum_{j\neq i}\frac{z_i}{z^2_i-z^2_j}\frac{\partial}{\partial z_i},\label{LB}
\end{eqnarray}
then we have the commutation relations
\begin{equation}
[O_L^{A,B}, O_E]=2O_L^{A,B}.
\end{equation}

It is known that the Hamiltonians of the $A_{N-1}$ and $B_{N}$-Calogero models are \cite{Sutherland,Yamamoto}
\begin{equation}\label{HA}
\hat H^A_C=\frac{1}{2}\sum_{i=1}^{N}(-\frac{\partial^2}{\partial z_i^2}+\omega^2 z_i^2)
+V^A,
\end{equation}
and
\begin{equation}\label{HB}
\hat H^B_C=\frac{1}{2}\sum_{i=1}^{N}(-\frac{\partial^2}{\partial z_i^2}+\omega^2 z_i^2)
+V^B,
\end{equation}
where the potentials $V^A$ and $V^B$ are given by $V^A=\frac{1}{2}\sum\limits_{i=1}^{N}
\sum\limits_{j\neq i}\frac{a(a-1)}{(z_i-z_j)^2}$, $V^B=\sum\limits_{i=1}^{N}\frac{b(b-1)}{2z_i^2}
+a(a-1)\sum\limits_{i=1}^{N}\sum\limits_{j\neq i}\frac{z^2_i+z^2_j}{(z^2_i-z_j^2)^2}$, 
respectively, the constants $a$ and $b$ are the coupling parameters, $\omega$ is the strength of the external harmonic well.

By performing a similarity transformation and removing the ground state from the Hamiltonian, we obtain
\begin{eqnarray}\label{HAC}
H^A_C&=&(\Psi^A_g)^{-1}(\hat H^A_C-E^A_g)\Psi^A_g\nonumber\\
&=&\sum_{i=1}^{N}(-\frac{1}{2}\frac{\partial^2}{\partial z_i^2}
+\omega z_i\frac{\partial}{\partial z_i})-a\sum_{i=1}^{N}\sum_{j\neq i}\frac{1}{z_i-z_j}\frac{\partial}{\partial z_i},
\end{eqnarray}
and
\begin{eqnarray}\label{HBC}
H^B_C&=&(\Psi^B_g)^{-1}(\hat H^B_C-E^B_g)\Psi^B_g\nonumber\\
&=&\sum_{i=1}^{N}(-\frac{1}{2}\frac{\partial^2}{\partial z_i^2}
-\frac{b}{z_i}\frac{\partial}{\partial z_i}+\omega z_i\frac{\partial}{\partial z_i})-2a\sum_{i=1}^{N}\sum_{j\neq i}\frac{z_i}{z_i^2-z_j^2}\frac{\partial}{\partial z_i},
\end{eqnarray}
where $\Psi^A_g$ and $\Psi^B_g$ are the ground state wave functions,  $E^A_g$ and $E^B_g$
are the ground state energies.

Then in terms of the Euler and Lassalle operators,
the Hamiltonians (\ref{HAC}) and (\ref{HBC})
can be rewritten in a unified fashion \cite{Nishino}
\begin{eqnarray}
H^{A,B}_C=\omega O_E-\frac{1}{2}O_L^{A,B},
\end{eqnarray}
where the parameters in $O_L^{A,B}$ (\ref{LA}) and (\ref{LB}) take
$\alpha=-1/a$, $\beta=2b$ and $\gamma=4a$.

Let us take the operators
\begin{eqnarray}\label{NOwalgebra}
\hat W_m^{r}&=&(\frac{O_E+N}{2})^{r-1}(O^A_L+V^A)^m\nonumber\\
&=&(\frac{1}{2}\sum_{i=1}^{N}\frac{\partial}{\partial z_i}z_i)^{r-1}
(\sum_{i=1}^{N}\frac{\partial^2}{\partial z_i^2}-\frac{2}{\alpha}\sum_{i=1}^{N}\sum_{j\neq i}\frac{\partial}{\partial z_i}\frac{1}{z_i-z_j})^m,
\end{eqnarray}
where $r\in \mathbb{Z}_+$, $m \in\mathbb{N}$ and we take $\alpha=\frac{4}{a(a-1)}$ in (\ref{LA}).
We then obtain the algebra
\begin{equation}\label{hatwalg}
[\hat W_{m_1}^{r_1}, \hat W_{m_2}^{r_2}]=(\sum_{k=0}^{r_2-1}C_{r_2-1}^{k}m_1^k
-\sum_{k=0}^{r_1-1}C_{r_1-1}^{k}m_2^k)\hat W_{m_1+m_2}^{r_1+r_2-1-k}.
\end{equation}
When particularized to the $r_1=r_2=2$ case in (\ref{hatwalg}), it gives the Witt algebra
\begin{equation}\label{walg}
[\hat W_{m_1}^{2}, \hat W_{m_2}^{2}]=(m_1-m_2)\hat W_{m_1+m_2}^{2}.
\end{equation}

By replacing the generators
$W_m^{r}\rightarrow \hat W_m^{r}= -\sum\limits_{n=-r+1}^{+\infty}\frac{m^{n+r-1}}{(n+r-1)!}W_n^{r}$
in the commutation relation (\ref{walgebra}),
after some simple calculation, it gives the commutation relation (\ref{hatwalg}).
Hence we also call (\ref{hatwalg}) the $W_{1+\infty}$ algebra.
It should be noted that not as the case of (\ref{walgebra}), (\ref{hatwalg}) contains
the operators $\hat W_{m}^{r}$ only for $m\ge 0$ and $r\ge 1$.
Precisely speaking (\ref{hatwalg}) is a subalgebra of the $W_{1+\infty}$ algebra.

By direct calculation of the $n$-commutator of (\ref{NOwalgebra}),
it gives the $W_{1+\infty}$ $n$-algebra
\begin{eqnarray}\label{NWnalgebra}
[\hat W_{m_1}^{r_1}, \hat W_{m_2}^{r_2}, \ldots, \hat W_{m_n}^{r_n}]
&=&(-1)^{\frac{n(n-1)}{2}}\epsilon_{1 2 \cdots n}^{i_1 i_2 \cdots i_n} \dsum_{\alpha_1=0}^{\beta_1}
\dsum_{\alpha_2=0}^{\beta_2}
\cdots \dsum_{\alpha_{n-1}=0}^{\beta_{n-1}}
C_{\beta_1}^{\alpha_1}C_{\beta_2}^{\alpha_2}\cdots C_{\beta_{n-1}}^{\alpha_{n-1}}
\nonumber\\
&&\cdot m_{i_2}^{\alpha_1}m_{i_3}^{\alpha_2}\cdots m_{i_n}^{\alpha_{n-1}}
\hat W_{m_1+m_2+\cdots +m_{n}}^{r_1+\cdots+r_n-(n-1)-\alpha_1-\cdots-\alpha_{n-1}}.
\end{eqnarray}
When $n$ is even, it is a generalized Lie algebra.

As the case of (\ref{Wnalgebra}), we can show that there are the following subalgebras
\begin{eqnarray}\label{sub2nalgebra}
[\hat W_{m_1}^{n+1}, \hat W_{m_2}^{n+1}, \ldots , \hat W_{m_{2n}}^{n+1}]=\dprod_{1\leqslant j<k\leqslant 2n}(m_k-m_j)
\hat W_{m_1+m_2+\cdots +m_{2n}}^{n+1},
\end{eqnarray}
and
\begin{eqnarray}\label{nullsubalg}
[\hat W_{m_1}^{n+1}, \ldots , \hat W_{m_{2n+1}}^{n+1}]=0,
\end{eqnarray}
where we take the scaled generators $\hat W_{m}^{n+1}\rightarrow \Lambda^{-\frac{1}{2n-1}}\hat W_{m}^{n+1}$, and the scaling coefficient
$\Lambda$ is given by
$\Lambda=(-1)^{n}\dsum_{(\alpha_1,\alpha_2,\cdots,\alpha_{2n-1})\in S_{2n-1} }C_{\beta_1}^{\alpha_1}
C_{\beta_2}^{\alpha_2}\cdots  C_{\beta_{2n-1}}^{\alpha_{2n-1}}
\epsilon_{1 2 \cdots {(2n-1)}}^{\alpha_1 \alpha_2 \cdots \alpha_{2n-1}}$
with $\beta_1=n$, $\beta_k=kn-\dsum_{i=1}^{k-1}\alpha_i,$ $2\leq k\leq 2n-1$.

It should be noted that the operators $\hat W_m^{r}$ are not the conserved operators, i.e.,
$[\hat W_m^{r}, \hat H_{C}^{A}]\neq 0$.
For the Calogero model (\ref{HA}) without the harmonic potential, its conserved operators
are constructed by the recursive definition \cite{Hikami}
\begin{eqnarray}
\widetilde{W}_{m}^r=\frac{1}{2(m+r)}[\sum_{i=1}^{N}z_i^2,\widetilde{W}_{m+2}^{r-1}],\ \ r\geq 2, m\geq -r+1,
\end{eqnarray}
and $\widetilde{W}_{m}^1=\sum_{j,k}(L^m)_{jk}$, the Lax operator $L_{jk}$ is given by
$L_{jk}=-{\bf i}\delta_{jk}\frac{\partial}{\partial z_j}+(1-\delta_{jk}) \frac{{\bf i}a}{z_j-z_k}$, ${\bf i}=\sqrt{-1}$.
These conserved operators constitute the $W_{1+\infty}$ algebra
\begin{eqnarray}\label{HWwalg}
[\widetilde{W}_{m_1}^{r_1}, \widetilde{W}_{m_2}^{r_2}]=[(r_2-1)m_1-(r_1-1)m_2]\widetilde{W}_{m_1+m_2}
^{r_1+r_2-2}+\cdots
\end{eqnarray}
where $\cdots$ indicates the lower-order terms corresponding to the quantum effect.
When $r_1=r_2=2$ in (\ref{HWwalg}), it gives the Witt algebra (\ref{walg}).
It should be pointed out that these conserved operators do not yield the closed $n$-algebra.

We have presented a realization of $W_{1+\infty}$ algebra in terms of the Euler and Lassalle operators
and the potential of the $A_{N-1}$-Calogero model.
Let us turn to introduce another realization
\begin{eqnarray}\label{NBwalgebra}
\check W_m^{r}&=&(\frac{O_E+N}{2})^{r-1}(O^B_L+V^B)^m\nonumber\\
&=&(\frac{1}{2}\sum_{i=1}^{N}\frac{\partial}{\partial z_i}z_i)^{r-1}
(\sum_{i=1}^{N}(\frac{\partial^2}{\partial z_i^2}+\frac{\partial}{\partial z_i}\frac{\beta}{z_i})+\gamma\sum_{i=1}^{N}\sum_{j\neq i}\frac{\partial}{\partial z_i}\frac{z_i}{z^2_i-z^2_j})^m,
\end{eqnarray}
where $r\in \mathbb{Z}_+$, $m \in\mathbb{N}$ and we take $\beta=\frac{1}{2}b(1-b)$, $\gamma=a(1-a)$ in (\ref{LB}).
Straightforward calculation shows that the operators (\ref{NBwalgebra}) also yield the
$W_{1+\infty}$ algebra (\ref{hatwalg}) and $n$-algebra (\ref{NWnalgebra}).

\section{$W_{1+\infty}$ constraints for the hermitian one-matrix model}

The partition function of the hermitian one-matrix model is
\begin{equation}
Z_{N}(\bt)=\int dM\ \ exp({\sum\limits_{k=0}^{\infty}t_k
\Tr M^k}),
\end{equation}
where $\bt=\{t_k|k\in \mathbb{N}\}$, $M$ is an $N\times N$ hermitian matrix and $dM$ is the Haar measure
\begin{equation}
dM=\prod_{i=1}^{N}dM_{ii}\prod_{i<j}d(\Re M_{ij})d(\Im M_{ij}),
\end{equation}
which is invariant under the gauge transformation $M\rightarrow UMU^{\dag}$,
and $U$ is a $U(N)$ matrix. In terms of the eigenvalues, the integral can be rewritten as
\begin{eqnarray}\label{HMM}
Z_N(\bt)=\int d^Nz \ \ \Delta(z)^2 exp(U(\bt)),
\end{eqnarray}
where $\Delta(z)=\prod\limits_{1\leq i<j\leq N}(z_i-z_j)$ and $U(\bt)=\sum\limits_{k=0}^{\infty}t_k\sum\limits_{i=1}^{N}z_i^k$.

An approach to derive the $W_{1+\infty}$ constraints for this matrix model was proposed in Ref.\cite{Itoyama}.
The following identity has been used there
\begin{equation}
\int d^Nz\ \ \Delta \cdot\mathcal{D}_{n,m}(exp(U)\Delta)
-(\mathcal{D}^{\dag}_{n,m}\Delta)\cdot exp(U)\Delta =0,
\end{equation}
where
$\mathcal{D}_{n,m}=\sum_{i=1}^{N}z_i^n\frac{\partial^m}{\partial z_i^m}$ and
$\mathcal{D}^{\dag}_{n,m}=(-1)^m\sum_{i=1}^{N}\frac{\partial^m}{\partial z_i^m}z_i^n, \ \ m, n\in \mathbb{Z}_{+}$,
whose Lie algebras are isomorphic to the $W_{1+\infty}$ algebra
(\ref{walgebra}).
The derived  $W_{1+\infty}$ constrains are
\begin{equation}\label{Iwconop}
\bar{\mathcal{W}}^s(P)Z_N(\bt)=0,\ \ s\geq 2,
\end{equation}
with
\begin{eqnarray}\label{wconop}
\bar{\mathcal{W}}^s(P)
&=&\frac{1}{s}:Q_s[j(P)+\frac{d}{dj}(P)]:_{-}
+(-1)^s\frac{1}{s}Q^{\dag}_s[\frac{d}{dj}(P)]\nonumber\\
&=&\sum_{n=-s+1}^{\infty}\bar{\mathcal{W}}^s_n P^{-n-s},
\end{eqnarray}
where $j(P):=\sum\limits_{k=1}^{\infty}kt_kP^{k-1}=\partial_PU(P)$, $\frac{d}{dj}(P):=\sum\limits_{k=0}^{\infty}P^{-k-1}\frac{\partial}{\partial t_{k}}$, the function $Q_s[f]:=(\frac{\partial}{\partial P}+f(P))^s\cdot 1$, $Q^{\dag}_s[f]:=(-\frac{\partial}{\partial P}+f(P))^s\cdot 1$,
the normal ordering $: \ \ :$ means we put the differential operator $j(P)$ before $\frac{d}{dj}(P)$
and the subscript $``- "$ is the projection to the negative powers of $P$.

When $s=2$, (\ref{wconop}) reduces to
\begin{equation}
\bar{\mathcal{W}}^2(P)=(\frac{d}{dj}(P))^2+(j\frac{d}{dj}(P))_{-}
=\sum_{n=-1}^{\infty}\bar{\mathcal{W}}^2_n P^{-n-2},
\end{equation}
where the operators $\bar{\mathcal{W}}^2_n$ are given by
\begin{equation}\label{virconop}
\bar{\mathcal{W}}^2_n=\sum_{k=0}^{\infty}kt_k\frac{\partial}{\partial t_{k+n}}+
\sum_{k=0}^{n}\frac{\partial}{\partial t_{k}}\frac{\partial}{\partial t_{n-k}},\ \ n\geq -1,
\end{equation}
which satisfy (\ref{walg}).
Thus we have the Virasoro constraints
\begin{equation}
\bar{\mathcal{W}}^2_nZ_{N}(\bt)=0.
\end{equation}

Taking $s=3$  in (\ref{wconop}), we obtain the constraints
\begin{equation}
\bar{\mathcal{W}}^3_nZ_{N}(\bt)=0,
\end{equation}
where
\begin{equation}
\bar{\mathcal{W}}^3_n=\sum_{k=0}^{\infty}kt_k\bar{\mathcal{W}}^2_{n+k}-(n+2)
\bar{\mathcal{W}}^2_{n},\ \ n\geq -2.
\end{equation}
It is difficult to write down the operators $\bar{\mathcal{W}}^s_n$
explicitly from (\ref{wconop}).
A conjecture is that the constraints (\ref{Iwconop}) with $s>2$
are reducible to the Virasoro constraints \cite{Itoyama}.

Let us focus on the partition function (\ref{HMM}) and insert the
operators (\ref{NOwalgebra}) under the integral as done in Ref.\cite{Nedelin}.
Then we have
\begin{equation}\label{defncon}
\int d^Nz \ \ \hat{W}_m^{r}(\Delta(z)^2 exp(U(\bt)))=0,\ \ r\in \mathbb{Z}_{+}, m\in \mathbb{N}.
\end{equation}

From the insertion of  $\hat{W}_0^{2}=\frac{1}{2}\sum\limits_{i=1}^{N}\frac{\partial}{\partial z_i}z_i$,
we obtain the identity
\begin{eqnarray}\label{w02id}
\int d^Nz \ \ \hat{W}_0^{2}(\Delta(z)^2 exp(U(\bt)))
&=&\int d^Nz \ \ \frac{1}{2}[\sum_{k=0}^{\infty}kt_k(\sum_{i=1}^{N} z_i^{k})+N^2](\Delta(z)^2 exp(U(\bt)))\nonumber\\
&=&\mathcal{O}_E\int d^Nz \ \ \Delta(z)^2 exp(U(\bt))=0,
\end{eqnarray}
where
\begin{eqnarray}\label{OE}
\mathcal{O}_E=\frac{1}{2}\sum_{k=0}^{\infty}kt_k\frac{\partial}{\partial t_{k}}
+\frac{1}{2}\frac{\partial^2}{\partial t_0^2}.
\end{eqnarray}
For the operator $\hat{W}_1^{1}=\sum\limits_{i=1}^{N}\frac{\partial^2}{\partial z_i^2}-\frac{2}{\alpha}\sum\limits_{i=1}^{N}\sum\limits_{j\neq i}\frac{\partial}{\partial
 z_i}\frac{1}{z_i-z_j}$, we have the action
\begin{eqnarray}\label{aw11id}
\hat{W}_1^{1}(\Delta(z)^2 exp(U(\bt)))
&=&[(2-\frac{2}{\alpha})\sum_{i=1}^{N}\sum_{j\neq i}\frac{1}{(z_i-z_j)^2}+(\frac{1}{\alpha}-1)\sum_{k=0}^{\infty}
k(k-1)t_k(\sum_{i=1}^{N} z_i^{k-2})\nonumber\\
&+&(2-\frac{1}{\alpha})\sum_{k=2}^{\infty}\sum_{l=0}^{k-2}kt_k (\sum_{i=1}^{N} z_i^l)(\sum_{j=1}^{N} z_j^{k-l-2})\nonumber\\
&+&\sum_{k,l=0}^{\infty}klt_kt_l(\sum_{i=1}^{N} z_i^{k+l-2})
](\Delta(z)^2 exp(U(\bt))).
\end{eqnarray}
Since the action of any differential operator with respect to the variables ${\bf t}$
on (\ref{HMM}) can not generate the term $\sum\limits_{i=1}^{N}\sum\limits_{j\neq i}\frac{1}{(z_i-z_j)^2}$
in (\ref{aw11id}), we insert the operator $\hat{W}_1^{1}$ with $\alpha=1$  under the integral.
Then we have
\begin{eqnarray}\label{w11id}
\int d^Nz \ \ \hat{W}_1^{1}(\Delta(z)^2 exp(U(\bt)))|_{\alpha=1}
&=&\int d^Nz \ \ [\sum_{k,l=0}^{\infty}klt_kt_l(\sum_{i=1}^{N} z_i^{k+l-2})\nonumber\\
&+&\sum_{k=2}^{\infty}\sum_{l=0}^{k-2}kt_k (\sum_{i=1}^{N} z_i^l)
(\sum_{j=1}^{N} z_j^{k-l-2})](\Delta(z)^2 exp(U(\bt)))\nonumber\\
&=&\mathcal{O}_L\int d^Nz \ \ \Delta(z)^2 exp(U(\bt))=0,
\end{eqnarray}
where
\begin{eqnarray}\label{OL}
\mathcal{O}_L=\sum_{k,l=0}^{\infty}klt_kt_l\frac{\partial}{\partial t_{k+l-2}}
+\sum_{k=2}^{\infty}\sum_{l=0}^{k-2}kt_k\frac{\partial}{\partial t_l}\frac{\partial}{\partial t_{k-l-2}}.
\end{eqnarray}

Note that the operators with the same expressions as (\ref{OE}) and (\ref{OL})
have also been presented for the Gaussian hermitian model \cite{Shakirov2009}.
Since the constraint operators $\mathcal{O}_E$ and $\mathcal{O}_L$
are associated with the (Euler) Lassalle operators,
here we call (\ref{w02id}) and (\ref{w11id}) the  Euler and Lassalle constraints, respectively.
The commutation relation between $\mathcal{O}_L$ and $\mathcal{O}_E$ is
\begin{equation}
[\mathcal{O}_L, \mathcal{O}_E]=-\mathcal{O}_L.
\end{equation}

By means of (\ref{w02id}) and (\ref{w11id}),
for the case of operators $\hat{W}_m^{r}$ with $\alpha=1$,
we may derive the constraints from (\ref{defncon})
\begin{equation}\label{wcon}
\mathcal{W}_m^{r}Z_N(\bt)=0,
\end{equation}
where the constraint operators are given by
\begin{equation}\label{wcongen}
\mathcal{W}_m^{r}=\mathcal{O}_L^m\mathcal{O}_E^{r-1}, \ \ r\in \mathbb{Z}_{+}, m\in \mathbb{N}.
\end{equation}

By direct calculation of the commutator of (\ref{wcongen}),
we obtain
\begin{equation}
[\mathcal{W}_{m_1}^{r_1},\mathcal{W}_{m_2}^{r_2}]
=(\sum_{k=0}^{r_1-1}C_{r_1-1}^{k}m_2^k
-\sum_{k=0}^{r_2-1}C_{r_2-1}^{k}m_1^k)\mathcal{W}_{m_1+m_2}^{r_1+r_2-1-k},
\end{equation}
which is isomorphic to the $W_{1+\infty}$ algebra (\ref{hatwalg}).

From (\ref{wcon}), we have the Virasoro constraints
\begin{equation}\label{vwcon}
\mathcal{W}_m^{2}Z_N(\bt)=0, \ \ m\in \mathbb{N},
\end{equation}
where the constraint operators are given by
\begin{equation}\label{vvcongen}
\mathcal{W}_m^{2}=
(\sum_{k,l=0}^{\infty}klt_kt_l\frac{\partial}{\partial t_{k+l-2}}
+\sum_{k=2}^{\infty}\sum_{l=0}^{k-2}kt_k\frac{\partial}{\partial t_l}\frac{\partial}{\partial t_{k-l-2}})^m
(\frac{1}{2}\sum_{k=0}^{\infty}kt_k\frac{\partial}{\partial t_{k}}
+\frac{1}{2}\frac{\partial^2}{\partial t_0^2}),
\end{equation}
which also satisfy (\ref{walg}).

Here we should like to draw attention to the fact that
both the Virasoro constraint operators (\ref{virconop}) and (\ref{vvcongen})
subject to the relation (\ref{walg}) and annihilate the partition function $Z_{N}(\bt)$ (\ref{HMM}).
However they are completely the different operators.
We have mentioned previously that the constraints (\ref{Iwconop}) for the hermitian
one-matrix model seem to be reducible to the Virasoro constraints.
Unlike that case, we observe that the $W_{1+\infty}$ constraints (\ref{wcon})
are indeed reducible to the Euler and Lassalle constraints.
An intriguing property of the constraint operators (\ref{wcongen}) is that they yield the closed $n$-algebra
\begin{eqnarray}\label{NWcongen}
[\mathcal{W}_{m_1}^{r_1}, \mathcal{W}_{m_2}^{r_2}, \ldots, \mathcal {W}_{m_n}^{r_n}]
&=&\epsilon_{1 2 \cdots n}^{i_1 i_2 \cdots i_n} \dsum_{\alpha_1=0}^{\beta_1}
\dsum_{\alpha_2=0}^{\beta_2}
\cdots \dsum_{\alpha_{n-1}=0}^{\beta_{n-1}}
C_{\beta_1}^{\alpha_1}C_{\beta_2}^{\alpha_2}\cdots C_{\beta_{n-1}}^{\alpha_{n-1}}
\nonumber\\
&&\cdot m_{i_2}^{\alpha_1}m_{i_3}^{\alpha_2}\cdots m_{i_n}^{\alpha_{n-1}}
\mathcal{W}_{m_1+m_2+\cdots +m_{n}}^{r_1+\cdots+r_n-(n-1)-\alpha_1-\cdots-\alpha_{n-1}},
\end{eqnarray}
which is isomorphic to the $W_{1+\infty}$ $n$-algebra (\ref{NWnalgebra}).

When particularized to the Virasoro constraint operators in (\ref{NWcongen}), it gives
the null $3$-algebra
\begin{equation}
[\mathcal{W}_{m_1}^{2}, \mathcal{W}_{m_2}^{2}, \mathcal{W}_{m_3}^{2}]=0.
\end{equation}
For the well-known Virasoro constraint operators (\ref{virconop}), it can be shown by direct
calculation that they do not yield any closed $n$-algebra.

Let us consider the insertions of the operators $\check W_m^{r}$ (\ref{NBwalgebra})
under the integral (\ref{HMM}).
For this case, we have
\begin{equation}\label{cWint}
\int d^Nz \ \ \check{W}_m^{r}(\Delta(z)^2 exp(U(\bt)))=0,\ \ r\in \mathbb{Z}_+, m \in\mathbb{N}.
\end{equation}
Although the integral will vanish under the insertions of $\check W_m^{r}$,
unfortunately, we can not derive the corresponding
$W_{1+\infty}$ constraints from (\ref{cWint}).

Making the changes of variables $z_i\rightarrow \sqrt{z_i}$
in the Euler operator $O_E$, Lassalle operator $O_{L}^{B}$ and
potential $V^{B}$, then we have $O_{E}\rightarrow 2 O_{E}$,
$O_{L}^{B}\rightarrow \widetilde{O}_{L}^{B}=
4\sum\limits_{i=1}^{N}z_i\frac{\partial^2}{\partial z_i^2}
+(2+2\beta)\sum\limits_{i=1}^{N}\frac{\partial}{\partial z_i}
+2\gamma\sum\limits_{i=1}^{N}\sum\limits_{j\neq i}
\frac{z_i}{z_i-z_j}\frac{\partial}{\partial z_i}$
and $V^{B}\rightarrow \widetilde{V}^B=\sum\limits_{i=1}^{N}\frac{b(b-1)}{2z_i}
+a(a-1)\sum\limits_{i=1}^{N}\sum\limits_{j\neq i}
\frac{z_i+z_j}{(z_i-z_j)^2}$.

Let us introduce the operators similar to (\ref{NBwalgebra})
\begin{eqnarray}\label{NBwalgebranew}
\breve W_m^{r}&=&(O_E+N)^{r-1}(\widetilde{O}^B_L+\widetilde{V}^B)^m\nonumber\\
&=&(\sum_{i=1}^{N}\frac{\partial}{\partial z_i}z_i)^{r-1}
(4\sum_{i=1}^{N}\frac{\partial^2}{\partial z_i^2}z_i+(2\beta-6)\sum_{i=1}^{N}\frac{\partial}{\partial z_i}-8\sum_{i=1}^{N}\sum_{j\neq i}\frac{\partial}{\partial z_i}\frac{z_i}{z_i-z_j})^m,
\end{eqnarray}
where we take $b=0$ and $\gamma=a(1-a)=-4$. It should be noted that the operators (\ref{NBwalgebranew}) also yield the
$W_{1+\infty}$ algebra (\ref{hatwalg}) and $n$-algebra (\ref{NWnalgebra}).

Inserting $\breve W_m^{r}$ (\ref{NBwalgebranew}) under the integral (\ref{HMM}), we may derive
\begin{equation}\label{bwcon}
\widetilde{\mathcal{W}}_m^{r}Z_N(\bt)=0,
\end{equation}
where the constraint operators are
\begin{equation}\label{bwcongen}
\widetilde{\mathcal{W}}_m^{r}=\widetilde{\mathcal{O}}_L^m(2\mathcal{O}_E)^{r-1}, \ \ r\in \mathbb{Z}_{+}, m\in \mathbb{N},
\end{equation}
the operator $\widetilde{\mathcal{O}}_L$ is given by
\begin{eqnarray}
\widetilde{\mathcal{O}}_L&=&(2\beta-2)\sum_{k=0}^{\infty}kt_k\frac{\partial}{\partial t_{k-1}}
+4\sum_{k=1}^{\infty}\sum_{l=0}^{k-1}kt_k\frac{\partial}{\partial t_l}\frac{\partial}{\partial t_{k-l-1}}
+4\sum_{k,l=0}^{\infty}klt_kt_l\frac{\partial}{\partial t_{k+l-1}},
\end{eqnarray}
which satisfies $\widetilde{\mathcal{O}}_LZ_N(\bt)=0$.

It can be shown that the constraint operators (\ref{bwcongen})
yield the same $W_{1+\infty}$ ($n$-)algebras as the cases of (\ref{wcongen}).

\section{Summary}
In terms of the Euler and Lassalle operators and the potentials of
the ($A_{N-1}$)$B_{N}$-Calogero models, we have presented the multi-variable
differential operator realizations of the $W_{1+\infty}$ algebra.
These operator realizations lead to the $W_{1+\infty}$ algebra and nontrivial $n$-algebra.
It should be noted that these operators are not the conserved operators for the Calogero model.
Based on the Lax operator of the Calogero model, the conserved operators of this system
which yield the $W_{1+\infty}$ algebra have been constructed in Ref.\cite{Hikami}.
However this type realization does not lead to the closed $n$-algebra.
Therefore the higher algebraic structures still deserve further study for the Calogero model.

We have reinvestigated the hermitian one-matrix model. From the insertions of our
realizations of the $W_{1+\infty}$ algebra under the integral, we have derived
the $W_{1+\infty}$ constraints, which are different from the constraints
presented in Ref.\cite{Itoyama}. The remarkable property of the derived constraint operators is that
they yield not only the $W_{1+\infty}$ algebra but also the closed $W_{1+\infty}$ $n$-algebra.
The higher algebraic structures should provide new insight into the matrix models.

\section *{Acknowledgment}

This work is supported by the National Natural Science Foundation
of China (Nos. 11875194, 11871350 and 11605096).


\end{document}